\documentclass[a4paper]{jpconf}
\usepackage{graphicx}
\usepackage{amssymb,amsmath,amsthm}
\usepackage{slashed}

\newcommand{\w}[1]{\mathbf{#1}}
\newcommand{\ws}[1]{\boldsymbol{#1}}
\newcommand{\be}{\begin{equation}}
\newcommand{\ee}{\end{equation}}
\newcommand{\bea}{\begin{eqnarray}}
\newcommand{\eea}{\end{eqnarray}}
\newcommand{\nn}{\nonumber}

\newcommand{\hajicek}{H\'aji\v{c}ek}

\font\tenscr=rsfs10 scaled1100
\font\sevenscr=rsfs7 % scaled \magstep1
\font\fivescr=rsfs5 % scaled \magstep1
\skewchar\tenscr='177
\skewchar\sevenscr='177
\skewchar\fivescr='177
\newfam\scrfam
\textfont\scrfam=\tenscr
\scriptfont\scrfam=\sevenscr
\scriptscriptfont\scrfam=\fivescr

\begin{document}
\title{A perspective on Black Hole Horizons from the Quantum Charged Particle}

\author{Jos\'e Luis Jaramillo}

\address{Universit\'e de Bretagne Occidentale, Brest, France}

\ead{jose-luis.jaramillo@univ-brest.fr}

\begin{abstract}
Black hole apparent horizons possess a natural notion of stability,
whose spectral characterization can be related to the problem
of the stationary quantum charged particle.
Such mathematical relation leads to an ``analyticity conjecture'' on the
dependence of the spectral properties on a complex ``fine-structure-constant''
parameter, that can reduce
the study of the spectrum of the (non-selfadjoint) MOTS-stability operator
to that of the (selfadjoint) Hamiltonian of the quantum 
charged particle. Moreover, this perspective might open an avenue 
to the spinorial treatment of apparent horizon (MOTS-)stability and to 
the introduction of semiclassical tools to explore some of the
qualitative aspects of this black hole spectral problem.

\end{abstract}

\section{The analogy: black hole apparent horizons and
 quantum charged particles}
Given a black hole apparent horizon section ${\cal S}$,
the systematic study of the spectral problem of the so-called
stability operator $L_{\cal S}$ \cite{AndMarSim05} of {\em marginally outer trapped surfaces} (MOTS), namely
\bea
\label{e:MOTS_stability_operator}
L_{\cal S} \psi \equiv \left[-\Delta  + 2 \Omega^a  D_a
- \left( |\Omega|^2
- D_a  \Omega^a  -\frac{1}{2}R_{\cal S} + G_{ab}k^a\ell^b \right)\right]\psi = \lambda \psi \ , 
\eea
has been proposed 
in \cite{Jaramillo:2013rda} as a methodology to explore aspects of the
black hole horizon geometry, possibly leading to insights into 
the black hole stability/instability problem (paraphrased,
%in \cite{Jaramillo:2013rda} 
following Kac's spectral discussion \cite{Kac66}, as
``can one hear the stability of a black hole horizon?'').

The general resolution of the MOTS-spectral problem in (\ref{e:MOTS_stability_operator}) 
represents a challenging task. In this sense, any hint relating this problem to
a better known and controlled problem is of clear relevance. This is 
precisely the context of Ref. \cite{Jaram14}, where a relation 
is presented between the MOTS-spectral problem and the study of the 
stationary states of a spin-$0$ quantum charged particle 
moving in the presence of  electric and magnetic fields.

To make this statement more precise, we start by
briefly describing the terms in the MOTS-stability operator $L_{\cal S}$.
Let us consider a $n$-dimensional spacetime with metric $g_{ab}$, associated Levi-Civita connection
$\nabla_a$ and Einstein curvature tensor $G_{ab}$. Let 
${\cal S}$ be a codimension-$2$ spacelike closed (compact without boundary) surface
with induced metric $q_{ab}$. $L_{\cal S}$ contains
both intrinsic and extrinsic geometry elements.
The intrinsic geometry ones associated with $q_{ab}$ are $D_a$, 
$\Delta=D^aD_a$ and $R_{\cal S}$, respectively
the Levi-Civita connection, the scalar Laplacian operator and the Ricci scalar
on ${\cal S}$. Let us also introduce the volume form $\epsilon_{a_1...a_{n-2}}$ with
associated measure $dS$.
Regarding the extrinsic geometry,
we introduce first two null vectors $\ell^a$ and $k^a$ spanning the 
normal bundle $T^\perp{\cal S}$, normalized as $k^a\ell_a=-1$. 
Then the \hajicek{} form $\Omega_a=-k^b{q_a}^c\nabla_c \ell_b$ encodes
part of the extrinsic geometry of ${\cal S}$, in particular 
providing a connection in the normal bundle. Physically, $\Omega_a$ represents 
an angular momentum density through the (Komar) expression 
$J=\int_{\cal S} \Omega_a \phi^a dS$, where $\phi^a$ is an axial Killing vector 
on ${\cal S}$. 
The term $G_{ab}k^a\ell^b$ is related to ambient spacetime dynamics through  
the Einstein equations $G_{ab}+\Lambda g_{ab}=8\pi T_{ab}$. Defining the (outgoing) expansion
on ${\cal S}$ as $\theta^{(\ell)}=q^{ab}\nabla_a\ell_b$, the MOTS condition on
${\cal S}$ is characterized as $\theta^{(\ell)}=0$, whereas the MOTS-stability
operator is defined \cite{AndMarSim05} through the variation of this expansion 
$L_{\cal S}\psi\equiv\delta_{-\psi k}\theta^{(\ell)}$, where $\psi$ is a function on ${\cal S}$
characterizing the surface deformation along $k^a$. Note that $L_{\cal S}$
is (formally) non-selfadjoint, namely due to the $2\; \Omega^a  D_a\psi$ term.

Given these elements, and as shown in \cite{Jaram14}, the MOTS-stability
operator is related to the Hamiltonian of the quantum charged
particle through the identifications
\bea
\label{e:MOTS_QuantumParticle_Analogy}
\Omega_a \leftrightarrow  \frac{ie}{\hbar c}A_a \ \ , \ \ R_{\cal S} 
\leftrightarrow \frac{4me}{\hbar^2}\phi
\ \ , \ \ G_{ab}k^a\ell^b \leftrightarrow -\frac{2m}{\hbar^2}V \ \ ,
\eea
where $A_a$ and $\phi$ correspond to the vector and scalar potentials
of given magnetic and electric fields on ${\cal S}$, whereas
$V$ represents an external mechanical potential. Indeed, $L_{\cal S}$ becomes
\bea
\label{e:quantum_Ham}
\frac{\hbar^2}{2m}L_{\cal S}\leftrightarrow \hat{H} &=& -\frac{\hbar^2}{2m}\Delta  + \frac{i\hbar e}{mc} A^a D_a
+ \frac{i\hbar e}{2mc} D_a A^a +  \frac{e^2}{2mc^2} A_a A^a 
+ e \phi + V \nn \\
&=& \frac{1}{2m}\left(-i\hbar D - \frac{e}{c} A\right)^2 
+ e\phi + V \ \ ,
\eea
where $\hat{H}$ is the Hamiltonian of the spin-$0$ quantum charged particle if the  
formal parameter $\hbar$ is interpreted as 
the Planck constant (over $2\pi$) and $m$ and $e$ as the mass and charge 
of the particle. The main points in \cite{Jaram14} are, first, to bring attention to the fact
that
the derivative and \hajicek{} terms in (\ref{e:MOTS_stability_operator})
can be combined as a perfect square into a single connection term, as follows
\bea
\label{e:MOTS_stability_operator_v2}
L_{\cal S} \psi = 
\left[-\left(D - \Omega\right)^2 +\frac{1}{2}R_{\cal S} 
- G_{ab}k^a\ell^b \right]\psi \ \ .
\eea
This underlines the role of $(D_a - \Omega_a)$ as the relevant connection in the
problem (actually a connection in the normal bundle $T^\perp {\cal S}$).
And secondly, pushing forward this simple remark, Ref. \cite{Jaram14} establishes
a structural connection between the MOTS-stability and quantum charged particle
problems beyond the purely formal analogy embodied in the correspondences 
(\ref{e:MOTS_QuantumParticle_Analogy}). Specifically, both problems possess
and Abelian gauge symmetry defined by the transformations
\be
\label{e:compared_gauge_transformations}
\begin{array}{ccclcl}
\hbox{Quantum Charged Particle:} & A_a &\to& A_a - D_a \sigma & , &  \psi \to e^{ie\sigma/(c\hbar)} \psi  \\
\hbox{MOTS:} &  \Omega_a &\to& \Omega_a - D_a \sigma & , &  \psi \to e^{-\sigma} \psi \ .\\ 
\end{array}
\ee
The first one defines the compact $U(1)$-gauge symmetry of electromagnetism, whereas the
second one defines a non-compact $\mathbb{R}^+$-gauge symmetry corresponding to the 
invariance of the MOTS-geometry under a rescaling of the null vectors (namely a
boost-transformation in the normal direction): $\ell'^a = e^{-\sigma}\ell^a, k^a = e^\sigma k^a $. 
In particular, this identifies the \hajicek{} form as the gauge potential of this  
$\mathbb{R}^+$-abelian symmetry. Moreover, as apparent in 
(\ref{e:MOTS_stability_operator_v2}) this gauge potential $\Omega_a$
(and therefore the black hole angular momentum) is introduced into the MOTS-problem via 
a {\em minimal coupling mechanism}, i.e. by substitution $D_a \rightarrow (D_a - \Omega_a)$
in the non-rotating problem. This is exactly the mechanism for switching on
magnetic (and electric) fields in the quantum particle. We can summarize
these gauge symmetries considerations in the following lemma \cite{Jaram14}:
 
\medskip

\noindent {\bf Lemma 1} {\em (MOTS-gauge transformations).}  \\
{\em Under the null normal rescaling $\ell'^a= f \ell^a$, $k'^a= f^{-1} k^a$, 
with $f>0$:
\begin{itemize}
\item[i)] The expansion and \hajicek{} form transform as:
%\bea
%\label{e:theta_Omega_transformations}
$\ \ \theta^{(\ell')}=f \theta^{(\ell)} \ $ and $\ \Omega'_a = \Omega_a + D_a(\mathrm{ln}f)$.
%\eea
\item[ii)] The MOTS-stability operator transforms covariantly:
%\bea
%\label{e:transformation_L_S}
$(L_{\cal S})' \psi  = f L_{\cal S} (f^{-1}\psi)$,
%\eea
where $(L_{\cal S})'\psi \equiv  \delta_{\psi (-k')} \theta^{(\ell')}$.
\item[iii)] The MOTS-eigenvalue problem is invariant under
the additional  eigenfunction transformation, $\psi'= f\psi$. That is,
%\bea
%\label{e:transformation_spectral_problem}
$L_{\cal S}\psi = \lambda \psi$ goes to $(L_{\cal S})' \psi'  = \lambda \psi'$.
%\eea
\end{itemize}
}
It is interesting to note that the gauge-connection $D^\Omega_a = D_a - \Omega_a$ 
is intimately related to the ``covariant derivative'' introduced in 
the GHP formalism \cite{gerochheld73:_ghp} to define a notion of 
``Fermi-Walker'' transport of a vector along the surface ${\cal S}$.
In particular, a vector $\gamma k^a$ is ``Fermi-transported'' along ${\cal S}$
if $(D_a - \Omega_a)\gamma=0$ (such transport is in general non-integrable 
on ${\cal S}$ and therefore dependent on the path). More generally
the GHP formalism, based on a choice of two null directions at each spacetime
point and naturally adapted to the study of codimension-$2$ surfaces,
could offer insights into the MOTS-stability problem through
its concepts of spin- and boost-weighted quantities (the latter directly
connected with the notion of MOTS-gauge symmetry).

The presented analogy between the MOTS-stability problem and the quantum
charged particle offers the possibility of transferring tools and concepts
from one problem to the other. Here we will comment: on the possibility
of addressing the MOTS-spectral problem by importing 
the expertise in the 
stationary spectrum of the quantum charged particle through an 
analytic continuation procedure, on the possibility of introducing 
semi-classical tools into the study of black horizon geometry and,
finally, on an approach to a spinorial formulation
of MOTS-stability by mimicking Pauli's and Dirac's steps in the
introduction of spinors in the quantum charged particle.   
We will conclude with a brainstorming list of possible directions of further 
study.

\section{MOTS-spectrum analyticity in the “fine structure constant” $\alpha$}
Let us introduce the ``fine structure constant'' $\alpha\equiv \frac{e^2}{\hbar c}$
and set
$\hbar\!\!=\!\!m\!\!=\!\!c\!\!=\!\!1$. %(so that in particular $\alpha= e^2$) 
We define the operator family $L[\sqrt{\alpha}]$
on the square root of this ``fine-structure-constant'' complex parameter $\alpha$, as
\bea
\label{e:L[alpha]}
L[\sqrt{\alpha}] &\equiv& -\frac{1}{2}(D -i\sqrt{\alpha}\Omega)^2 
- \frac{\alpha}{4}R_{\cal S} - \frac{1}{2}G_{ab}k^a\ell^b \nn \\ %= - \frac{1}{2}\Delta\\ 
%= -\frac{1}{2}\Delta  \\  
&=&  -\frac{1}{2}\Delta + i\sqrt{\alpha}(\Omega\cdot D
+ \frac{1}{2}D\cdot \Omega) + \frac{\alpha}{2}|\Omega|^2 
- \frac{\alpha}{4}R_{\cal S} - \frac{1}{2}G(k,\ell) \ , 
\eea
so that the quantum charged particle corresponds to a real positive $\alpha = 1$ 
(normalized as
$e^2=1$), whereas 
(half) the MOTS-stability operator corresponds to a real negative fine structure 
constant $\alpha=-1$. More precisely, we make branch choices 
 $\hat{H}=L[\sqrt{\alpha}=1]$ and $L_{\cal S}/2=L[\sqrt{\alpha}=-i]$. 

The operator family $L[\sqrt{\alpha}]$ provides an ``analytical continuation'' in 
$\sqrt{\alpha}$ from the self-adjoint quantum charged particle Hamiltonian $\hat{H}$ to the
non-selfadjoint MOTS-stability operator $L_{\cal S}$. This naturally raises the following question:
{\em can we recover the MOTS-spectrum ($\alpha=-1$) as an 
analytic extension of the quantum charged particle spectrum ($\alpha=1$) self-adjoint problem?}.
If this is indeed possible, we can perform the following ``self-adjoint trick'' to
reduce the resolution of the non-selfadjoint spectral problem associated with $L_{\cal S}$ 
to a self-adjoint one: first
multiply formally $\Omega_a$ by $\sqrt{\alpha}$, then rotate $\sqrt{\alpha}$ in the complex plane 
to $i\sqrt{\alpha}$ to produce a self-adjoint operator whose spectral problem can be 
explicitly solved using standard techniques, and finally perform a ``back rotation''
$\sqrt{\alpha}\to \frac{1}{i}\sqrt{\alpha}$ in the explicitly obtained eigenvalues.

Answering the question above in its full generality
sets a difficult problem in %the setting of 
perturbation theory of 
linear operators \cite{Kato80}. 
%Our particular MOTS scenario
%is free of three pathologies:
%{\em i)} ${\cal S}$ has no boundaries 
%(it is closed), {\em ii)} the functions in $L[\sqrt{\alpha}]$ 
%can be taken as regular as needed, and {\em iii)}, potential topological issues
%associated to the underlying $U(1)$ or $ \mathbb{R}^+$-fibre bundle are absent
%since such bundle is trivial by construction (there exists a global
%connection $\Omega_a$). 
We formulate the following conjecture \cite{Jaram14}
as an open problem:

\medskip
\noindent {\bf Analyticity Conjecture.}  
{\em Given an orientable closed surface ${\cal S}$
and the one-parameter family of operators $L[\sqrt{\alpha}]$ defined 
in (\ref{e:L[alpha]}),  (in the complex $\sqrt{\alpha}$), 
the MOTS-spectrum 
($\sqrt{\alpha} = -i$) can be recovered as an ``analytic continuation''
of the quantum charged particle spectrum ($\sqrt{\alpha} = 1$).
} 
\medskip

In order to provide some support on the validity of this conjecture, we consider
an explicit example that we can fully solve and that presents the essential 
qualitative features to be expected in the generic case. This simple example
provides an analogue in our context to the Landau levels 
for the quantum charged particle in a constant magnetic field in $\mathbb{R}^3$.
We take as ${\cal S}$ a topological sphere $S^2$ endowed
with the round metric $q_{ab}=r^2(d\theta^2 + \sin^2 \theta d\varphi^2)$.
On $S^2$, the Hodge decomposition of the \hajicek{} form is 
$\Omega_a = {\epsilon_a}^b D_b \omega + D_a \zeta$. We make the simplest
non-trivial choice $\omega= \sqrt{\alpha} \cos \theta, \zeta =0$ 
(with $\sqrt{\alpha}\in\mathbb{R}$)
 that leads to
$\Omega = \sqrt{\alpha} \sin^2 \theta d\varphi$. Finally, we assume vacuum
($G_{ab} + \Lambda g_{ab}=8\pi T_{ab}=0$). 
With these choices, the relevant terms in (\ref{e:MOTS_stability_operator}) are
\bea
R_{\cal S}=\frac{2}{r^2} \ \ , \ \ 2\Omega^aD_a\psi = \frac{2\sqrt{\alpha}}{r^2}\partial_\varphi \psi
 \ \ , \ \ |\Omega|^2 = \frac{\alpha}{r^2}\sin^2\theta \ \ , \ \ D^a\Omega_a = 0 \ \ , \ \
G_{ab}k^a\ell^b = \Lambda \ .
\eea
We insert this in Eq. (\ref{e:MOTS_stability_operator}), separate variables as
$\Psi = S(\theta)e^{im\varphi}$ and write $\lambda=\lambda_{\mathrm R}
+ i \lambda_{\mathrm I}$. The imaginary part of the equation
produces $\lambda_{\mathrm I}= 2 \sqrt{\alpha} m/r^2$, 
whereas the real part ($x=\cos\theta$) is 
\bea
\label{e:spheroidalharmonics_prolate}
\frac{d}{dx}\left((1-x^2)\frac{d}{dx}\right)S + \left(\left((\lambda_{\mathrm{R}} + \Lambda)r^2 + \alpha 
-1\right) -\alpha x^2 - \frac{m^2}{1-x^2}\right)S=0 \ .
\eea
Solutions to this  equation are given \cite{AbramowitzStegun64}  
by the (prolate) spheroidal harmonics 
 $S_{\ell m}(\sqrt{\alpha}, cos\theta)$, with  
 values $\lambda_{\ell m}(\sqrt{\alpha})= (\lambda_{\mathrm{R}} + \Lambda)r^2 + \alpha -1$.
The solutions of the MOTS-problem are therefore
\bea
\label{e:MOTS_prolatespheroidal}
\lambda = \frac{\lambda_{\ell m}(\sqrt{\alpha}) + 1 - \alpha}{r^2} - \Lambda + 
i \frac{2\sqrt{\alpha}m}{r^2} \ \ , \ \
\psi_{\ell m}(\theta,\phi) =   S_{\ell m}(\sqrt{\alpha},\cos\theta) e^{im\varphi} \ .
\eea
To study the corresponding quantum charged particle problem, we 
make $\sqrt{\alpha}\to i\sqrt{\alpha}$ in $\Omega_a$. Denoting the new
eigenvalues as $\bar{\lambda}$ and the eigenfunctions by $\bar{\psi}$ and repeating the steps above, we
first obtain $\lambda_{\mathrm I}=0$ (the operator is now self-adjoint),
whereas the equation on $x$ is now
\bea
\label{e:spheroidalharmonics_oblate}
\frac{d}{dx}\left((1-x^2)\frac{d}{dx}\right)S + \left(\left((\bar{\lambda}_{\mathrm{R}} + \Lambda)r^2 + 
2\sqrt{\alpha}m - \alpha 
-1\right) +\alpha x^2 - \frac{m^2}{1-x^2}\right)S=0 \ \ .
\eea
This is now the  equation for the (oblate) spheroidal harmonics 
\cite{AbramowitzStegun64} with solutions $\bar{S}_{\ell m}(\sqrt{\alpha}, cos\theta)=
S_{\ell m}(i\sqrt{\alpha}, cos\theta)$ and 
where the corresponding 
$\bar{\lambda}_{\ell m}(\sqrt{\alpha})= (\bar{\lambda}_{\mathrm{R}} + \Lambda)r^2 
+ 2\sqrt{\alpha} m - \alpha -1$ satisfy
$\bar{\lambda}_{\ell m}(\sqrt{\alpha})=\lambda_{\ell m}(i\sqrt{\alpha})$ (note
that $\lambda_{\ell m}(i\sqrt{\alpha})$ are real numbers \cite{AbramowitzStegun64}). 
The solution is now
\bea
\label{e:QCP_oblatespheroidal}
\bar{\lambda} = \frac{\lambda_{\ell m}(i\sqrt{\alpha}) + 1 + \alpha -2\sqrt{\alpha}m}{r^2} - 
\Lambda  \ \ , \ \
\bar{\psi}_{\ell m}(\theta,\phi) =   S_{\ell m}(i\sqrt{\alpha},\cos\theta) e^{im\varphi} \ .
\eea
This is in agreement with the ``analyticity conjecture'' formulated above,
since both eigenvalues and eigenfunctions in the solution (\ref{e:MOTS_prolatespheroidal}) 
to the non-selfadjoint problem are indeed recovered from the solutions 
of the self-adjoint case
(\ref{e:QCP_oblatespheroidal}), when applying back the shift
 $\sqrt{\alpha}\to \frac{1}{i}\sqrt{\alpha}$.

\section{Semi-classical tools in the MOTS-spectral problem. 
Towards an action principle for MOTS-stability}
The possibility of reducing the spectral problem %of the non-selfadjoint
of $L_{\cal S}$ to that of a selfadjoint  operator, admitting
in addition the  interpretation of a quantum Hamiltonian, opens the 
possibility of considering a particular semi-classical approach to the study of
$L_{\cal S}$.
Inverting the quantization rule $p_i \to -i D_i$ (with $\hbar=1$), the
Hamiltonian $\hat{H}(\sqrt{\alpha})\equiv L[\sqrt{\alpha}]$ leads to
the classical Hamiltonian function
\bea
\label{e:classical_Ham}
H_{\mathrm{cl}}[\sqrt{\alpha}](x,p) = (p - \sqrt{\alpha}\Omega)^2 
+ \frac{1}{2}R_{\cal S} - G(k,\ell) \ .
\eea
Then, we can use $H_{\mathrm{cl}}[\sqrt{\alpha}](x,p)$ as the starting
point to obtain approximate solutions to the spectral problem of $\hat{H}(\sqrt{\alpha})$
by employing semi-classical tools such as the WKB techniques \cite{Berry83}. Under the assumption 
of the validity of the ``analyticity conjecture'', the relevant expressions 
approximating the MOTS-problem would be then obtained by evaluating
 $\sqrt{\alpha}\to -i$.

A relevant question is: can we
define an action from which the MOTS-stability problem
can be derived variationally? The action for a (complex) scalar field coupled
to an electromagnetic field provides the answer in the self-adjoint case, but 
its straightforward application to $L_{\cal S}$ fails. We have not
been able to find a real action from which $L_{\cal S}$ emerges variationally.
On the other hand, considering a complex scalar $\psi$ on ${\cal S}$ we can 
introduce the following complex action
\bea
\label{e:MOTS_action}
S = \int_{\cal S} dS \left( q^{ab}(D_a + \Omega_a)\psi^*(D_b - \Omega_b)\psi 
+ \left(\frac{1}{2}R_{\cal S} - G_{ab}k^a\ell^b\right)\psi^*\psi - F_\psi \psi^* - F_{\psi^*}\psi 
    \right) \ .
\eea
Variating independently $\psi^*$ and $\psi$, the corresponding Euler-Lagrange 
equations are
\bea
L_{\cal S}\psi &=& \left(-(D-\Omega)^2  + \frac{1}{2}R_{\cal S} - G_{ab}k^a\ell^b
\right)\psi =  F_\psi \nn\\
 (L_{\cal S})^\dagger\psi^* &=& \left(-(D+\Omega)^2 + \frac{1}{2}R_{\cal S} - G_{ab}k^a\ell^b
\right)\psi^*  =  F_{\psi^*} \ .
\eea
Choosing $F_\psi=\lambda\psi$, with $\lambda$ a parameter to be determined,
we recover the eigenvalue problem (\ref{e:MOTS_stability_operator}).
One recovers not only the elliptic problem for 
$L_{\cal S}$, related to the ingoing variation of the outgoing expansion
$L_{\cal S}\psi = \delta_{-k\psi}\theta^{(\ell)}$,
but also the problem $(L_{\cal S})^\dagger\psi^*$ related to 
the outgoing variation of the ingoing expansion: $(L_{\cal S})^\dagger\psi^* 
= \delta_{-\ell\psi^*}\theta^{(k)} - \kappa^{(\psi^*\ell)}\theta^{(k)}$.
This suggests that the relevant object in this context is actually a 
two-component vector
related to second variations of the element of area. ``Diagonal'' 
second variations
would correspond to Raychaudhuri equations, whereas the ``non-diagonal'' 
ones are 
expressed in terms of the MOTS-stability operator and its adjoint.

\section{Spinors and MOTS-stability}
Having a spinor (first-order) characterization of MOTS-stability
would present various potential interests: i) inner boundary conditions 
for elliptic problems with a horizon (e.g. in Witten's proof
of mass positivity, approaches to Penrose-like inequalities...),
ii) reduction of the spectral problem to that of a first-order operator and, 
more generally, iii)  natural setting for studying MOTS-stability when studying 
spinorial fields propagating on a black hole spacetime background.

The analogy of the MOTS-stability operator with the Hamiltonian of the
quantum charged particle provides a natural way for introducing spinors 
by mimicking Pauli's and Dirac's approaches to particles' spin. 
%spin degree of freedom in the 
Let us first introduce
some notation. Considering a tetrad ${e^a}_i$, such that $g_{ab}{e^a}_i {e^b}_{j}=\eta_{ij}$,
assuming a spin-structure we introduce
gamma-matrices $\ws{\gamma}^a ={e^a}_i\ws{\gamma}^i$ acting on spinors $\w{\Psi}$ and satisfying
the Clifford algebra $\{\ws{\gamma}^a,\ws{\gamma}^b\}=2g^{ab}\w{1}$.
In the Riemannian case of the MOTS ${\cal S}$, we consider 
$\{\ws{\gamma}^a,\ws{\gamma}^b\}=2q^{ab}\w{1}$. The derivative connection on spinors is  
$\w{D}_a\ws{\Psi}=\left(D_a 
+\frac{1}{8}\omega_a^{ij}[\ws{\gamma}_i,\ws{\gamma}_j]\right)\ws{\Psi}$, where $\omega_a^{ij}$ is the 
standard spin-connection associated with ${e^a}_i$. 

A straightforward approach to recast  MOTS-stability in terms of a first-order condition, 
would be to take the ``square-root'' of $L_{\cal S}$ in the same spirit in which 
the square root of the Klein-Gordon equation of a field of mass $m$, 
namely $((i\hbar)^2\square + m^2c^2)\Psi=0$, is the
Dirac equation $(i\hbar \gamma^iD_i + mc)\w{\Psi}=0$. However, this 
does not work for $L_{\cal S}$, due to the non-constancy of the corresponding ``mass terms''.
As an alternative, we consider the second-order Pauli equation 
starting from the following key remark: 
the Laplacian (in Euclidean $\mathbb{R}^3$ and acting on
spinors) can be written in two ways as follows ($\ws{\sigma}^i$ are Pauli matrices, the gamma matrices
in this setting)
\bea
\label{e:2Laplacians}
\Delta=D^i D_i=\left(\ws{\sigma}^i D_i\right)^2 \ \ , \ \ \{\ws{\sigma}^i,\ws{\sigma}^j\}= 
2\delta^{ij}\ws{1} \ (\hbox{Clifford relations}) \ .
\eea 
%\be
%\label{e:Pauli_matrices}
%\ws{\sigma}^1= \left(\begin{array}{cc} 0& 1\\ 1&0\end{array}\right) \ , \
%\ws{\sigma}^2=\left(\begin{array}{cc} 0& -i\\ i&0\end{array}\right) \ , \
%\ws{\sigma}^3=\left(\begin{array}{cc} 1& 0\\ 0&-1\end{array}\right) 
%\ee
The crucial point is that introducing the magnetic
vector potential through minimal coupling actually depends on 
the starting version we choose for the Laplacian.
In particular, the spin-magnetic field coupling 
term $\frac{\hbar e}{2mc}\ws{\sigma}^iB_i$, with giromagnetic factor $2$, 
is recovered when performing
$\Delta=\left(\ws{\sigma}^i D_i\right)^2 \to \left(\ws{\sigma}^i (D_i - ie A_i)\right)^2$.
In the curved MOTS case, Eq. (\ref{e:2Laplacians}) is generalized through (an adaptation of)
the  Lichnerowicz-Weitzenb\"ock formula (we denote $F^\Omega_{ab}=D_a\Omega_b-D_b\Omega_a$)
%(i\slashed{D}_\Omega)^2 = 
\bea
\label{e:Lichnerowicz-Weitzenbock}
\left(i\ws{\gamma}^a (\w{D}_a - \Omega_a) \right)^2
= -(D_a - \Omega)^2 + \frac{1}{4}R_{\cal S} 
+ \frac{1}{4}[\ws{\gamma}^a,\ws{\gamma}^b]F^\Omega_{ab} \ .
\eea
The MOTS-stability operator can then be written as
\bea
L_{\cal S} = %(i\slashed{D}_\Omega)^2 
\left(i\ws{\gamma}^a (\w{D}_a - \Omega_a) \right)^2
+ \frac{1}{4}R_{\cal S} - \frac{1}{4}[\ws{\gamma}^a,\ws{\gamma}^b]F^\Omega_{ab} - G_{ab}k^a\ell^b \ .
\eea
From the Pauli equation perspective, the $F^\Omega_{ab}$ term modifies the ``spin-magnetic
field'' coupling term
correcting the ``giromagnetic factor'' and then indicating that, from this point of view,
the MOTS is a composite object. More importantly, although this is still a second-order
operator, it indicates the relevance of the first-order operator defined from
the codimension-$2$ Sen connection, 
$i\ws{\gamma}^a (\w{D}_a - \Omega_a)$, and its related spectral problem.
Finally, Pauli's second-order equation can be obtained from the
Dirac first-order one by taking the (non-relativistic) limit $c\to 0$. This remark
indicates a formal path to define a first-order spinorial operator from which 
the MOTS-stability operator $L_{\cal S}$ can be recovered through a limit procedure.
This will be presented elsewhere \cite{Jaramillo:2015}.

\section{Perspectives}
The analogy between stable MOTS and quantum particles 
(stable MOTS behave as charged particles with ``negative fine-structure constant'')
provides the seed for a research program relying on the transfer of
knowledge between black hole and quantum charged particle physics.

We list some possible directions of research: i)  
self-adjoint ``shortcut'' to the spectral MOTS-problem through the analyticity conjecture;  
ii) applications to the MOTS-stability of Kerr;
iii)  MOTS-spectrum statistics
and applications of a ``$L_{\cal S}$-spectral zeta function $\zeta_{_{L_{\cal S}}}(s)$'' constructed by analytic
continuation from  the $\zeta_{_{L[\sqrt{\alpha}]}}(s)$ of the selfadjoint $L[\sqrt{\alpha}]$; 
iv) semiclassical 
(and dynamical-systems) tools in the study of the MOTS-stability operator,
in particular to construct approximate explicit expressions for MOTS-stability in Kerr,
as well as generic ``high-eigenvalue'' asymptotics; %and link to quantum billiards.
v) spinor reformulation of MOTS-stability and applications to 
inner boundary conditions in elliptic problems relevant for the ``mass problem'' in 
General Relativity (positivity, Penrose-like inequalities, quasi-local gravitational mass); 
vi) variational derivation of MOTS-stability from an action functional
%: Action from Wess-Zumino term in a Chern-Simons action, 
and relation to Ginzburg-Landau theory on closed Riemannian manifolds; 
%(link to Seiberg-Witten theory in the $\mathrm{dim}({\cal S})=4$ self-dual case)...
vii) higher-dimensional black hole horizons, with
richer topologies and field content through the Hodge-decomposition %of the \hajicek{}
$\Omega = d\alpha + \delta \beta + \gamma$ (with $\gamma$ harmonic).
Finally, we mention other complementary issues: 
gauge-invariant expression of the charged particle ground state from Donsker-Varadhan
theory \cite{Jaram14},
possibility of a MOTS ``Aharonov-Bohm-like effect'', 
signature of quasi-normal modes/superradiance, ``second-quantization''-approach
(motivated by a ``many-particle'' treatment of the Schr\"odinger equation), 
study of effective  MOTS-deformation ``degrees of freedom'' and possible 
statistical-mechanics/thermodynamical properties of MOTS.

\medskip

\noindent {\em Acknowledgments}. I thank J.M.M. Senovilla and J.A. Valiente-Kroon for discussions.
\medskip

%%%%%%%%%%%%%%%%%%%%%%%%%%%%%%%%%%%%%%%%%%%%%%%%%%%%%%

%%%%%%%%%%%%%%%%%%%%%%%%%%%%%%%%%%%%%%%%%%%%%%%%%%%%%%

\end{document}